# Postulates and Experimental Features in Maxwell's Electromagnetic Equations

## A. Orefice

Universita' di Milano - DI.PRO.VE. - Via Celoria, 2 - 20133 - Milano (ITALY)
adriano.orefice@unimi.it

**ABSTRACT** The Maxwell integral equations expressing Ampere's and Faraday's laws are shown to be affected by heavy physical approximations. The usual deduction from them, moreover, of the corresponding set of differential Maxwell equations is based, in general, on a wrong use of the Stokes theorem. The equivalence, therefore, between the two sets of equations must be reconsidered. Interesting conclusions may be drawn from the comparison between the experimental set-ups of the Faraday's law and of the Aharonov-Bohm effect.

## I. INTRODUCTION

Standard textbooks of classical electromagnetism sometimes **[1]** axiomatically postulate, from the very outset, Maxwell's electromagnetic equations in their differential form

$$\nabla \times \boldsymbol{E} = -\frac{1}{c}\frac{\partial \boldsymbol{B}}{\partial t} \tag{1}$$

$$\nabla \times \boldsymbol{B} = \frac{4\pi \boldsymbol{J}}{c} + \frac{1}{c}\frac{\partial \boldsymbol{D}}{\partial t}. \tag{2}$$

The Lorentz invariance of these equations, together with the wave equation they provide, allows, in principle, a coherent and causal description of any possible electromagnetic feature.

There exist, however, many phenomena which appear to be much more directly described by the integral (i.e. macroscopic) form of Maxwell's equations:

$$\oint \boldsymbol{E}\cdot d\boldsymbol{l} = -\frac{1}{c}\int_S \frac{\partial \boldsymbol{B}}{\partial t}\cdot d\boldsymbol{S} \tag{3}$$

$$\oint \boldsymbol{H}\cdot d\boldsymbol{l} = \frac{4\pi}{c}\int_S \boldsymbol{J}\cdot d\boldsymbol{S} + \frac{1}{c}\int_S \frac{\partial \boldsymbol{D}}{\partial t}\cdot d\boldsymbol{S} \tag{4}$$

where $d\boldsymbol{S} = \boldsymbol{n}\, dS$, and $\boldsymbol{n}$ is a unit vector normal to the element $dS$ of the arbitrary surface $S$ around which the line integral is computed. It is in such an integral form, indeed, that both Faraday's and Ampere's laws were originally formulated.

The use of eqs.(3) and (4) would cause, of course, no conceptual trouble if their complete equivalence with eqs.(1) and (2) could be assumed to hold.

It is however our aim to show that such an equivalence, in its usual form, is not granted, both for mathematical and physical reasons.

## II. THE ROLE OF THE STOKES THEOREM

Let us consider a physical situation where a magnetic field, strictly confined within a limited region (e.g. a long and narrow solenoid) is allowed to vary in time. It is then customary to say that, if we look at an arbitrary loop encircling the region where $\mathbf{B} \neq 0$ (and placed even very far from it, in a zone where the magnetic field is and remains negligible) an electromotive force (*emf*) is induced along such a loop, according to the equation

$$\oint \mathbf{E} \cdot d\mathbf{l} = \int_S \nabla \times \mathbf{E} \cdot d\mathbf{S} = -\frac{1}{c} \int_S \frac{\partial \mathbf{B}}{\partial t} \cdot d\mathbf{S} \equiv -\frac{1}{c} \frac{d\Phi_B}{dt} \tag{5}$$

where $S$ is any surface covering the loop and $\Phi_B$ is the magnetic flux through it. Now, the first two terms of eq.(5) simply represent Stokes' theorem, expressing a geometrical identity when referred to stationary situations. In a time-dependent situation such as the one considered here, however, since eq.(5) connects two distinct and separate regions of space (the loop and the encircled surface $S$) we must expect that the field variation occurring in the solenoid launches an electromagnetic signal, raising in the surrounding space a time-varying electric field

$$\mathbf{E} = -\frac{1}{c} \frac{\partial \mathbf{A}}{\partial t} \tag{6}$$

with a vector potential $\mathbf{A}(r,t)$ provided by the expression

$$\mathbf{A}(\mathbf{r},t) = \frac{1}{c} \int_V dV \frac{\mathbf{J}(\mathbf{r}', t - \frac{|\mathbf{r} - \mathbf{r}'|}{c})}{|\mathbf{r} - \mathbf{r}'|} \tag{7}$$

where $\mathbf{J}$ represents the current density distribution in a volume $V$ containing the solenoid, and $\mathbf{r}'$ is evaluated within the volume element $dV$. Clearly enough, the volume $V$ (where the time-varying current density $\mathbf{J}$ is distributed) the surface $S$ (through which the flux is computed) and the loop (along which the *emf* is induced) constitute separate and distant regions of space, between which eq.(5) neither establishes nor predicts any kind of propagation, since each one of its terms is computed *at the same time*. Eq.(5) surreptitiously introduces, in other words, an *instantaneous* physical connection. Not surprisingly, the overall result

$$emf = -\frac{1}{c}\frac{d\Phi_B}{dt} \qquad (8)$$

is evidently *non-local*: it asserts, in fact **[2]**, that the electromotive force is *simultaneous* to the magnetic flux variation causing it, wherever it may be generated. In other words, cause and effect (although arbitrarily distant) are declared to be simultaneous, thus implying an *instantaneous*, and therefore *unphysical*, transmission of energy and information.

## III. THE ROLE OF EXPERIMENTAL EVIDENCE

Strangely enough eq.(8), *applied to an arbitrary line*, is exactly the form of Faraday's law usually presented as directly provided by the experimental evidence. All standard textbooks, in fact, agree on the fact that it is easily verified by interrupting a conducting loop in a point and inserting an electrometer between its free extremities. As we have seen, however, eq.(8), because of its instantaneous transmission of information, cannot not reflect a physical law. It could represent, at most, a first approximation, where the propagation time is neglected: an excusable mistake for experiments performed in strictly limited spaces, but a quite inexplicable conceptual oversight.

It may be observed that eq.(5) is generally written backwards **[2,3]**, thus claiming to obtain Maxwell's differential equation (1) from the integral one (3).

The same *non-equivalence* proof, of course, may be shown to hold between eqs.(2) and (4), connecting the electric field variations with the displacement currents.

## IV. THE CASE OF THE AHARONOV-BOHM EFFECT

We recall now that Aharonov and Bohm **[4]** have proposed a famous experiment where a coherent electron beam, launched from an electron source *ES*, is split into two parts, successively recombined at a point *P* after having travelled along two equal, but mutually distant, paths (which we shall call 1 and 2). The ensemble of the two paths from *ES* to *P* is assumed to form a symmetric loop around a large region of space containing at its centre a very limited zone (for instance, a long and tightly wound solenoid normal to the plane of the loop) where a strong, constant magnetic field is established. Outside this zone and along the electron paths the magnetic field is assumed to be completely negligible.

Aharonov and Bohm considered the superposition of the wave functions of the two recombined electron beams, showing that it leads to a (Young-like) interference pattern at the point P presenting, with respect to the case of no magnetic field at all, a phase shift given by the expression

$$\frac{e}{c\hbar}\oint \boldsymbol{A}\cdot d\boldsymbol{l} = \frac{e}{c\hbar}\int_S \nabla\times\boldsymbol{A}\cdot d\boldsymbol{S} = \frac{e}{c\hbar}\int_S \boldsymbol{B}\cdot d\boldsymbol{S} \equiv \frac{e}{c\hbar}\Phi_B \qquad (9)$$

where the first integral is computed along the closed circuit composed by the paths 1 and 2, the second integral is due to Stokes' theorem, and $\Phi_B$ is the magnetic flux across any surface $S$ enclosed by such a circuit. The phase shift predicted by eq.(9) was experimentally observed by Chambers **[5]**, as well as by many other researchers.

Although Aharonov and Bohm attribute this effect to the action of the magnetic potential **A** (which would therefore turn out to be endowed with a real, and not only conventional, physical nature), it is most often interpreted **[6]** as a *non-local* action-at-a-distance between the electrons and the distant magnetic field.

A strong analogy is clearly present between eq.(9), concerning the (quantum) Aharonov-Bohm effect, and eq.(5), concerning the (classical) Faraday law. Both cases refer, indeed, to a large loop, encircling a wide, fieldless region where the magnetic field is different from zero only inside a narrow zone. In both cases, moreover, an important role is played by the magnetic flux: an ambiguous physical quantity (independent from the field distribution over the surface to which it refers) which almost unavoidably suggests a non-local phenomenology, whenever such a surface is of finite size and the field distribution is inhomogeneous.

There are, however, at least two fundamental (and strictly interconnected) differences between the two cases.

In the first place, eq.(9) refers to a *stationary situation*, where the use of Stokes' theorem is certainly allowed.

In the second place the non-locality of the Aharonov-Bohm effect (as well as that of the so-called "EPR paradox" **[7]**) would occur without transmission of energy and/or information, i.e. without any physical contradiction, *as long as it is referred to a steady state situation*.

Clearly enough, however, no time modulation of the magnetic flux $\Phi_B$ may be instantaneously revealed by the observed time variation of the (arbitrarily far away) interferential phase shift, since it could be employed for an unphysical instantaneous transmission of information.

This is strictly related, of course, to the fact that a *simple-minded* term by term time derivative of eq.(9) would lead, by means of eq.(6), to the unacceptable eq.(5).

Both Faraday's law (5) and a *wrong use* of the (perfectly correct) Aharonov-Bohm eq.(9) lead therefore to analogous forms of instantaneous information transmission.

**V. CONCLUSION**

The ambiguous character of Maxwell's equations (experimental inductions or axiomatic postulates?) was already pointed out in previous papers (see, for instance, Refs. **[8-11]**).

We stress in the present work one more element of ambiguity. If we try, in fact, to obtain Maxwell's differential equations (1) and (2) - as is done in most standard textbooks - from the integral ones (3) and (4) applied to arbitrarily large regions of space, we start from *incorrect* experimental statements, and make, moreover, an *incorrect* use of Stokes' theorem.

A possible way out from this situation could be found by taking eqs.(3)-(8) in the limit $S \to 0$ (i.e., in practice, by rejecting Maxwell's integral equations in their most general form), since any non-locality would disappear when restricting the considered area to a single point.

Another - substantially equivalent - way out could be provided by simply assuming the differential Maxwell equations (1) and (2) as axiomatic postulates: a quite inelegant approach, indeed, for an experimental science.

Both Faraday's and Ampere's integral laws must be seen, in any case, as roughly approximate statements, largely unable to describe, in general, macroscopic physical events.